\documentstyle[prl,aps]{revtex}
\begin{document}
\draft
\title{  Cosmic Fluctuations and  Dark Matter
from   Scalar Field Oscillations }
\author{Craig J. Hogan}
\address{Astronomy and Physics Departments FM-20, University of Washington,
    Seattle, WA 98195}
\date{\today}
\maketitle
\def\msol{{\,\rm M_{\odot}}}
\begin{abstract}
Scale-invariant    fluctuations and   cold
dark matter could originate from
two different modes  of
a single scalar field, fluctuations from
 massless Goldstone oscillations    and    matter from massive
Higgs modes. Matching the
 fluctuations and dark matter density   observed  requires a
heavy scale ($\phi_0\approx 10^{16}$GeV) for the potential
minimum  and an extremely small self coupling
($\lambda\approx 10^{-83}$). Mode coupling causes the dark matter to
form in lumps  with
nonnegligible velocities, leading to early collapse of dense
dark matter
``miniclusters'' and halos on the scale of compact dwarf galaxies.
\end{abstract}
\section{Introduction}
It is generally acknowledged that new physics is required to explain
the origin of cosmic structure, both in the cosmic background
radiation and in the galaxy distribution. At some level, all suggestions at
present
require activity of a new  scalar field.  The favorite hypothesis is that
quantum
fluctuations in the fields driving cosmic inflation create scale-invariant
fluctuations\cite{jim,rocky,salopek,dick}. If inflation leaves
 behind a
 universe which is too smooth,
similar structures can be introduced later by large-scale classical motions
of scalar
fields\cite{physica,zeldovich,vilenkin,alex,turok,spergel,stebbins,pen}.
New physics is
also required to produce the apparent prevalence of nonbaryonic,
nonrelativistic dark
matter  in the universe. Among the many possibilities are scalar and
pseudoscalar bosons
such as the axion,  which condense  into nonrelativistic dark matter during
vacuum phase
transitions\cite{john,larry,dine}. Usually however the fields considered as
dark
 matter
candidates have little to do with the sources of cosmic fluctuations.

In this paper I explore the
  possibility that the two unresolved phenomena---
large scale fluctuations and dark matter--- might arise from two different
modes
of oscillation of a single global scalar field.
 I use a
simple Mexican-hat model potential to describe the generic
(topology-independent) classical dynamical behavior of multicomponent
scalars in an expanding universe, in which classical modes are excited
by the Kibble mechanism. I argue that in general the Goldstone modes of
the field   produce   scale-invariant fluctuations and the Higgs modes
become cold dark matter.
Combining these functions into one field is not merely economical and tidy,
but makes a physical difference: during the
early stages of oscillation when the Higgs amplitudes are large, the modes
couple to
each other, leading to additional isocurvature fluctuations and peculiar
velocities
in the Higgs matter.

The behavior is determined quantitatively by  two standard parameters
describing
the
shape of the potential $V(\phi)$. The width of the hat,
$\phi_0$, controls the strength of the scale-invariant gravitational
perturbations; the self-coupling,
$\lambda$, controls the amount of dark matter.
Fixing  $\phi_0$ to match the
COBE/DMR fluctuation  amplitude on large scales\cite{cobe},
  the present density of  dark matter  requires a tiny self-coupling parameter,
$\lambda\approx10^{-83}$.
With these parameters, mode coupling causes the Higgs matter to condense in
lumps
of astrophysically interesting sizes and velocities.

There is little new here apart from the particular way a lot of
familiar ideas are spliced together from  many
other scenarios  which create cosmic structure with  active scalars.
The Goldstone-mode part of the model discussed here  corresponds to
massive global cosmic strings with mass per length $\approx\phi_0^2$--- even
though the strings are much thicker, and much less dense, than the
usual situation considered with potentials of central height $V_0\approx
\phi_0^4$.
The self-ordering dynamics  create
scale-invariant fluctuations  even in theories with no topological
defects\cite{alex}; simulations of this process\cite{pen} have however
excluded Higgs
degrees of freedom. The Higgs-mode condensate is a standard element in
inflation
theory\cite{rocky,kofman}, and resembles     the direct-condensation process
for
forming  cosmic axions\cite{rocky,john,larry,dine} (although here it is the
radial
and not the axial degrees of  freedom that provide the harmonic potential
to condense
in). The coupling of Higgs  with Goldstone modes resembles   radiation of
axions
from cosmic strings\cite{axionstrings1,axionstrings2}. The formation of the
isocurvature fluctuations is by  a process similar to that envisioned for
``axion
miniclusters''\cite{miniclusters,igor}, except that the scale is now large
enough to be
more astrophysically interesting. The same condensation process was also
considered in a
late-phase-transition model\cite{press}  with still weaker coupling;   that
scenario
however included no Goldstone modes. Models with  weaker coupling still
(i.e., Higgs masses $m\equiv 2\lambda \phi_0^2<H_0\approx 10^{-32}$eV) and
$\phi_0\approx  m_P(H_0/m)$ (and no Goldstone modes, which would cause
disastrous
fluctuations in this situation)
 can produce a  cosmological constant
\cite{weinberg,carroll,fukugita,ratra,frieman}. This work   shows that
in some supergravity models,   exponentially small masses arise
naturally from anomalies, so there is at least a plausible context for the
parameters
 proposed here.

\section{Higgs and Goldstone Modes}

For definiteness  consider the behavior of a complex classical scalar
field  described by Lagrangian density\cite{rocky}
\begin{equation}
L=\partial_\mu\phi\partial^\mu \phi/2-V(\phi),
\end{equation}
with a
potential
of the familiar form
\begin{equation}
V(\phi)= ({\lambda/ 4})(\phi^2-\phi_0^2)^2,
\end{equation}
assuming as usual $c=\hbar=1$.
The  potential has the form of a Mexican hat of height at center
$
V_0=\lambda \phi_0^4/4
$
and a set of degenerate minima forming a circle at
$|\phi|=\phi_0$, characterized by an internal phase angle $\theta$.
The field obeys the
evolution equation
\begin{equation}
\ddot \phi+3H\dot\phi-\nabla^2\phi+{\partial V/\partial \phi}
=0,
\end{equation}
where $H=\dot a/a$ and $a$ denotes the cosmic scale factor.
As this simple theory has no dissipative couplings, there is no
``reheating'' in the
theory\cite{rocky,kofman}, and energy  losses are all adiabatic.

The system supports two kinds of oscillations.  The first are
``classical Goldstone modes,'' corresponding to motion within the
 circle of minima. Quantum mechanically these modes are massless
 Goldstone bosons;
the classical modes are oscillations in which the gradient term
$\nabla^2\phi$ plays the role of a restoring force which tries to
correct misalignments in $\phi$. In these modes  the field is not
perfectly aligned but has spatial variations in
$\theta$; the misalignments  propagate
in space  at unit
velocity, with a characteristic frequency determined by the
wavelength.

The second type of oscillations are ``classical Higgs modes,''
corresponding to harmonic motion in the hat's radial direction.
The characteristic   frequency is
\begin{equation}
\omega^2=V''(\phi_0)=2m^2=2\lambda\phi_0^2
\end{equation}
where $m$ is the mass of the Higgs particle. The zero momentum modes
are spatially uniform and do not
propagate.

Both types of modes carry energy with distinctive
equations of state. The density and pressure are\cite{rocky}
\begin{equation}
\rho_\phi=\dot\phi^2/2 + V(\phi)+ (\nabla\phi)^2 /2
\end{equation}
\begin{equation}
p_\phi=\dot\phi^2/2 - V(\phi)- (\nabla\phi)^2 /6
\end{equation}
where the gradient term contributes an anisotropic pressure
in the direction of $\nabla\phi$.
A
static uniform $\phi$  field yields the familiar inflationary (de
Sitter, steady state, cosmological constant) equation of state,
$p=-\rho=-V$;
a
changing $\phi$ contributes an ultra-stiff component,
$p=\rho=\dot\phi^2$; and
a stationary
spatial gradient contributes  $p=-{1\over
3}\rho=- (\nabla\phi)^2/6$  (Which incidentally in spite
of carrying energy,
 has zero Newtonian
gravity; a universe made of such material mimics an open universe
even with zero space curvature).

 On timescales   comparable to the
oscillation period, the pressure and density of the modes
 fluctuates between these various
extreme equations of state.
On timescales long compared to an oscillation period,
the
equations of state of the two types of  modes are harmonic time
averages of these expressions over the oscillation.   The
Goldstone modes have $\langle V\rangle =0$
so their equation of state averages $\dot\phi^2/2$ and
$ \langle(\nabla\phi)^2\rangle $, with the latter multiplied times
$+{1\over 2}$ and
$-{1\over 6}$ for the density and pressure respectively, to produce
simple relativistic matter with
$p=\rho/3$. In the the zero-momentum
Higgs modes the gradient vanishes and the
average of $V$ and $\dot\phi^2/2$ produces a
cancellation, yielding pressureless matter  $p=0$; this phenomenon is
familiar from the formation of a cosmological axion condensate.
In both cases the density and pressure are proportional to the
squared amplitude of the oscillations.
\section{Fluctuations and Dark Matter}

Now consider the evolution of the classical  field in an expanding
universe. The effective potential at high temperatures is as usual
driven to have a minimum at $\phi=0$, which is the
initial condition of the system apart from small fluctuations. For
$T<2\phi_0$, this minimum disappears, so that the classical system
rolls down from the center of the hat to someplace on the circle of
minima.  This process in general excites both types of modes, but in
different ways and at different times.

The Goldstone modes are excited by the same Kibble mechanism
responsible for the formation of cosmic strings.
(In fact, for the specific potential used here strings actually form
as well; but the Goldstone oscillations are in addition to any such
topological defects). Widely separated portions of the universe choose
independently the value of
$\theta$ they relax to; thus  any random initial conditions of the
field naturally produce large-amplitude (in $\theta$) spatial
gradients in
$\phi$ on all scales.  These gradients are approximately
preserved,
frozen in comoving coordinates, on each comoving scale as
long as the wavelength exceeds $H^{-1}$; they excite propagating
Goldstone modes of unit amplitude on each scale when the wavelength
comes within
$H^{-1}$.

The expansion rate is
related to the cosmic density $\rho$ by the Friedmann equation
$
H^2=\rho/m_P^2
$
where $m_P\equiv (3\hbar c/ 8\pi G)^{1/2}=4.5 \times 10^{18}$GeV is the
Planck mass. The Goldstone modes on each scale contribute  density
fluctuations when they cross the horizon of the order of their
fluctuating density. Since they always move locally at the speed of light,
the  gradient is determined by the expansion rate $H$,
\begin{equation}
\delta\rho = \rho_\phi \approx (\nabla\phi)^2
\approx \delta\theta^2 \phi_0^2H^2=\rho(\phi_0/m_P)^2  \delta
\theta^2.
\end{equation}
The modes on each scale are initially excited  with $  \delta \theta
\approx 1$. (The amplitude decreases by Hubble damping   after they
come within the horizon and oscillations begin, leading to a
stochastic background of [practically undetectable] relativistic  Goldstone
waves.) This process thus produces
approximately scale-free fluctuations with amplitude
$\delta\rho/\rho\approx (\phi_0/m_P)^2$.
 Matching the COBE/DMR amplitude requires
$(\phi_0/m_P)^2
\approx 10^{-5}$, or $\phi_0\approx 10^{16}$GeV.
 Detailed simulations, and exact solutions for the (nearly Gaussian) case
of many $\phi$
components, confirm  the nearly scale-invariant spectrum, and  provide
the precise normalization of CBR anisotropy to matter fluctuations
\cite{turok,spergel,pen}.  In general, gravitational effects of strings or
textures, if
any,    are up to logarithmic factors comparable to (and additional to) the
 Goldstone
effect.

The radial Higgs mode is also excited in this system, just by
starting at the top of the hat (at the origin), rolling off and
overshooting the minimum. The natural
timescale for initial relaxation to produce the zero-momentum  mode
is just the oscillation time
$\omega^{-1}$.
Even if the rolling starts very early ($T\approx \phi_0$), the
oscillations as such do not start until $\omega\approx H$.
(This is also when the Higgs and Goldstone modes decouple
from each other; after this the coupling is weak because the Higgs amplitude
is $<<\phi_0$ and the Higgs
frequency no longer matches the $\delta\theta=1$ Goldstone modes.) The
temperature
$T_{\omega_H}$ when this happens can be estimated by setting
$
\omega^2=2\lambda\phi_0^2
$
equal to
$
H^2=a_sT_{\omega_H}^4/m_P^2,
$
yielding
\begin{equation}
{T_{\omega_H}/ m_P}=(2/a_s)^{1/4}\lambda^{1/4}(\phi_0/m_P)^{1/2}.
\end{equation}
(This has assumed radiation domination $\rho_{rad}=a_sT^4$,
where $a_s=\pi^2 N_{eff}/15$ and $N_{eff}$ is the number of
effective photon degrees of freedom).
At this time, the density in the Higgs modes, like the Goldstone modes
at  all times as they enter the horizon, is
\begin{equation}
(\rho_{\phi}/ \rho_{rad})_{\omega H}
\approx
V_0/a_sT_{\omega H}^4=(\phi_0/m_P)^2\approx 10^{-5}
\end{equation}
The subsequent behavior is very different however. Although subsequent
excitation of Higgs modes is very inefficient, these initial excitations,
like those of the Goldstone modes,
have their amplitudes reduced only by the expansion. However, as the
Higgs correspond to pressureless matter, they do not lose energy as
quickly as the relativistic matter:
$\rho_{\phi}\propto \rho_{rad}
/T$. They contribute a  component
of pressureless dark matter today whose ratio to the radiation density
at the present temperature $ T_0 \approx 5\times 10^{-32} m_P$ is then
\begin{equation}
\rho_{\phi}/ \rho_{rad}
\approx(\phi_0/m_P)^2 (T_{\omega_H}/T_0)
\approx(\phi_0/m_P)^{2.5} (m_P/T_0)\lambda^{1/4}.
\end{equation}
Let $\Omega_\phi$ denote the fraction of critical density today in
the form of   cosmic cold dark matter in the Higgs oscillations, and
$\Omega_{rad}=4\times 10^{-5}h^{-2}$ the fraction in radiation.
A reasonable density requires a tiny coupling;
$\Omega_\phi h^2\approx 4\times 10^{20}\lambda^{1/4}$,
$
\lambda\approx 10^{-83}(\Omega_\phi h^2)^4$,
and a tiny Higgs mass,
$m=\sqrt{2}\lambda^{1/2}\phi_0\approx 6\times 10^{-17}(\Omega_\phi h^2)^2$eV.
The characteristic energy scale of the
unbroken vacuum is $ V_0^{1/4}=\phi_0\lambda^{1/4}\approx 20
(\Omega_\phi h^2)$keV.
The formation of the Higgs condensate occurs at
 redshift $z_{\omega H}\approx (\rho_\phi/ \rho_{rad})_0
(\phi_0/m_P)^{-2}\approx 2.5 \times 10^{9}(\Omega_\phi h^2)$,   at a
temperature
 $T_{\omega H}\approx 7\times 10^{9 }(\Omega_\phi h^2)$K,
 or about 600$(\Omega_\phi
h^2)$keV.
The Compton wavelength
of the Higgs particles  is
 $2\pi/m\approx 2\pi/H\approx 2\times 10^{12 }(\Omega_\phi
h^2)^{-2}$cm, corresponding to a characteristic oscillation period
of about 60$(\Omega_\phi h^2)^{-2}$sec.

The Higgs
is not excited  in a spatially uniform   zero momentum mode.
The initial mixing with the Goldstone modes ensures fractional
variations of the order of unity in the rolloff time, since the
spatial gradients in the field accelerate  rolloff in some places and
retard  it in others, leading to spatial variations in
the phases   and
  initial epochs for the oscillations. The amplitude of the
Higgs oscillation therefore
  fluctuates spatially, with about unit fractional amplitude on the
scale of the Hubble length at $T_{\omega H}$. This variation
 leads  to dark matter forming in coherent lumps,
 with the Higgs matter
having coherent peculiar velocity of the order of unity on this scale.
The lumps are created  with a distribution of dark matter masses roughly in
the range
$[1{\rm\ to}\ (2\pi)^3]\times  V_0/m^{3}
\approx[2{\rm\ to}\ 550]
 (\Omega_\phi h^2)^{-2}\msol$, with
$100 (\Omega_\phi h^2)^{-2}\msol$   a typical value.
These ``compensated isocurvature fluctuations''\cite{peebles} in dark
matter density form
in addition to the Goldstone modes considered already. Since the lumps are
laid down
with no
 large-scale
correlations (according to original hypothesis of the Kibble domain
formation), the
{\it isocurvature} fluctuation spectrum corresponds  to white noise on
larger scales.

\section{Astrophysical Consequences}

Isocurvature fluctuations   grow at a rate of the order of
$(\rho_\phi/m_P^2)^{-1/2}$--- slower than $H$  until
  the epoch of equal
matter and radiation densities $t_{eq}$, at the usual rate $H$ thereafter.
 Eventually the perturbations
grow to be nonlinear and collapse into bound dark matter
``miniclusters''\cite{miniclusters,igor} in virial equilibrium.
A spherical model\cite{igor}
estimates the density of the virialized system after collapse,
$\rho_F=140\Delta^3(\Delta +1)\rho_{eq}$, where $\rho_{eq}=3\times
10^{-16}(\Omega_0 h^2)^4{\rm g \ cm^{-3}}$ is the matter density at $t_{eq}$,
and $\Delta$ is the initial fractional  overdensity.
For the first miniclusters, on the scale of the Higgs lumps, $\Delta$
 is of the order of unity.
Linear
 initial fluctuations on   scales larger than this
have  the white noise spectrum of rms
fluctuations in spheres of mean mass $M$,
\begin{equation}
\Delta_{rms}(M)\approx [  M/10^2 (\Omega_\phi
h^2)^{-2}\msol]^{-1/2}\Omega_\phi/\Omega_0.
\end{equation}
These   fluctuations cause hierarchical clustering   earlier than
with standard CDM,
starting with the minicluster formation and proceeding to larger scales.
Applying the
spherical model for each mass scale in the hierarchy of clustering, the
dark matter
forms into  virialized systems   which    have
a
  virial radius
\begin{equation}
 10^{16}{\rm\  cm} [M/10^2 (\Omega_\phi
h^2)^{-2}\msol]^{5/6}
[(\Omega_0 h^2)(\Omega_\phi h^2)^5]^{-1/3},
\end{equation}
 and
a virial velocity $\approx \sqrt{GM/R}$
\begin{equation}
\approx 10{\rm\ km\ sec^{-1}}
[M/10^2 (\Omega_\phi
h^2)^{-2}\msol]^{1/12}
(\Omega_0/\Omega_\phi)^{1/6},
\end{equation}
a characteristic velocity    high enough
 that
 clusters can accrete
 baryons  at high redshift.
 The computed velocity dispersion approximately matches   that
of  the  densest
dark halos yet found,  compact dwarf
galaxies\cite{dwarf1,dwarf2,dwarf3,dwarf4}. Although halos of this
dispersion extend to lower masses, and higher density, than in standard CDM, on
larger scales the fluctuations are dominated by the
 Goldstone modes and the
predictions are almost the same.

The Higgs matter is not perfectly cold, since the same coupling that creates
the   lumps also
produces  spatial gradients in Higgs mode phase,
corresponding to particle  peculiar velocities.
Although still  cold compared to massive neutrinos, the relict particle
velocities for the Higgs matter are much larger than for the usual CDM
candidates
such as WIMPs or axions. The velocity
dispersion is of the order of
$( \rho_{rad}/\rho_{\phi})(\phi_0/m_P)^2=3{\rm km/sec}$ at
$t_{eq}$. Since this is comparable  to the virial velocity of the miniclusters,
they  form with nonsingular cores comparable in size to the virial
radius\cite{moore}.

Another constraint is that dynamics cannot increase  the
 classical particle phase space density
\cite{tremaine} .
The initial condensation creates mildly relativistic
($\gamma\approx 1$) particles with mass density $\approx
\phi_0^2m^2$; thus there are $\phi_0^2m$ particles per volume, spread out over
a
wavenumber volume of order $m^3$.   Although the fine-grained phase space
density of
the particles is extremely high, once orbits cross and coherence is lost the
effective coarse-grained phase density is
$\approx
\phi_0^2/m^2$, leading to an upper limit on the density of subsequently
formed systems
of velocity dispersion
$v$:  $\rho<\phi_0^2m^2v^3$, which is  about
$
4\times 10^{-15}(v/10{\rm km\ sec^{-1}})^3
(\Omega_\phi h^2)^4\ {\rm g\ cm^{-3}}
$
for the parameters determined above.

If any of the  original miniclusters survive they might be detected by
gravitational microlensing of   quasars.
The Einstein radius at the Hubble distance is\cite{refsdal,ramesh}
\begin{equation}
R_E=2\sqrt{GM/H}= 8\times 10^{17}  h^{-1/2} (M/10^2 \msol)^{1/2} {\rm \ \ cm}.
\end{equation}
The smallest miniclusters are themselves smaller than this, so individual
miniclusters are compact
enough
to amplify distant objects significantly,
allowing a variety of observational probes\cite{refsdal,ramesh}. The
incidence  of such
events however depends on the number of surviving miniclusters, which is
difficult to
compute.

\acknowledgments
This work was supported by NASA grants NAGW-2569 and NAG5-2793 at the
University of Washington.

 \end{document}